\documentclass[iop]{emulateapj}

\newcommand{\cha}{\textit{Chandra }}
\def\xmm{{XMM-{\it Newton\/}}}

\def\lu{{ erg s$^{-1}$}}
\def\swi{{{\it Swift}-BAT}\/}

\def\lu{{ erg s$^{-1}$}}
\def\flu{{ erg s$^{-1}$} cm$^{-2}$}

\def\nustar{{\it NuSTAR}}

\shorttitle{X-ray spectral properties of seven heavily obscured Seyfert 2 galaxies}
\shortauthors{Marchesi et al.}

\usepackage{natbib}
\usepackage{float}
\usepackage{color}
\usepackage{graphicx}
\usepackage{gensymb}
\usepackage{enumitem}
\usepackage{longtable}
\begin{document}

\slugcomment{Accepted for publication in the Astrophysical Journal on January 14, 2017}
\title{X-ray spectral properties of seven heavily obscured Seyfert 2 galaxies}

\author{S. Marchesi\altaffilmark{1}, M. Ajello\altaffilmark{1}, A. Comastri\altaffilmark{2}, G. Cusumano\altaffilmark{3}, V. La Parola\altaffilmark{3}, A. Segreto\altaffilmark{3}}

\altaffiltext{1}{Department of Physics \& Astronomy, Clemson University, Clemson, SC 29634, USA}
\altaffiltext{2}{INAF--Osservatorio Astronomico di Bologna, via Ranzani 1, 40127 Bologna, Italy}
\altaffiltext{3}{INAF - Istituto di Astrofisica Spaziale e Fisica Cosmica, Via U. La Malfa 153, I-90146 Palermo, Italy}

\begin{abstract}
We present the combined \cha and \swi\ spectral analysis of seven Seyfert 2 galaxies selected from the \swi\ 100-month catalog. We selected nearby ($z\leq$0.03) sources lacking of a ROSAT counterpart and never previously observed with \cha in the 0.3--10\,keV energy range, and targeted these objects with 10\,ks \cha ACIS-S observations. The X-ray spectral fitting over the 0.3--150\,keV energy range allows us to determine that all the objects are significantly obscured, having N$_{\rm H}\geq10^{23}$\,cm$^{-2}$ at a $>$99\% confidence level. 
Moreover, one to three sources are candidate Compton thick Active Galactic Nuclei (CT-AGN), i.e., have N$_{\rm H}\geq10^{24}$\,cm$^{-2}$.
We also test the recent ``spectral curvature'' method developed by \citet{koss16} to find candidate CT-AGN, finding a good agreement between our results and their predictions.
Since the selection criteria we adopted have been effective in detecting highly obscured AGN, further observations of these and other Seyfert 2 galaxies selected from the \swi\ 100-month catalog will allow us to create a statistically significant sample of highly obscured AGN, therefore better understanding the physics of the obscuration processes.
\end{abstract}


\section{Introduction}
The Cosmic X-ray Background (CXB), i.e., the diffused X-ray emission observed in the energy range going from 1 to $\sim$200--300\,keV, is mainly caused by unobscured and obscured Active Galactic Nuclei \citep[AGN; e.g.,][]{alexander03,gilli07,treister09}.
However, while the unobscured AGN population, responsible for the CXB emission at $<$10 keV, has been almost completely detected, at the present day only $\sim$30\% of the CXB at its peak \citep[$\sim$30\,keV,][]{ajello08a} has been directly detected, thanks to several surveys with the \nustar\ telescope \citep{aird15,civano15,harrison15,mullaney15}. Moreover, while theoretically Compton thick (CT-) AGN, i.e., sources with absorbing column density N$_{\rm H}\geq10^{24}$\,cm$^{-2}$, are expected to be numerous \citep[see, e.g.,][]{risaliti99} and responsible for $\sim$10--25\,\% of the CXB emission at its peak, the fraction of CT-AGN detected with respect to the total number of AGN is so far $\sim$5\% \citep{comastri04,dellaceca08}. 

For AGN with column densities $\leq$10$^{25}$\,cm$^{-2}$, a fraction of direct nuclear emission can be observed at $\geq10$\,keV; for larger column densities, instead, the only observable component in the same energy range is the scattered, indirect one \citep[see, e.g., ][]{matt99,yaqoob10}. Consequently, a complete census of the CT-AGN population requires deep observations above 10\,keV. \citet{burlon11} analyzed the 15--55\,keV spectral properties of a complete sample of AGN at $z<$0.1 detected by \swi, founding that the observed fraction of CT-AGN in this sample is $\sim$5\%. However, in the same work it was pointed out that even above 10\,keV the AGN emission is strongly suppressed by absorption and Compton scattering. Therefore, the $\sim$5\% fraction of obscured AGN observed in most works in literature could not be intrinsic, but could actually being related to an observational bias. In fact, the modelling of this selection effect in \citet{burlon11} suggests that the intrinsic CT-AGN population accounts for about 20\% of the whole AGN population. A similar intrinsic fraction of Compton thick sources has been computed also by \citet{brightman11} and \citet{vasudevan13}.

While detecting CT-AGN remains a challenging task, in the last ten years several works were able to detect at least a part of these extremely obscured objects. Since at $z>$1 the $>$10\,keV energy range is observable in the 0.5--10\,keV band, deep-field surveys with \cha and \xmm\ have been able to detect CT-AGN at $z>$1 \citep[see, e.g., ][]{georgantopoulos13,vignali14,buchner15,lanzuisi15}. A further step in detecting and analyzing candidate CT-AGN was the launch of \nustar, which allowed to study single sources with unprecedented statistics in the 5--50\,keV band \citep[see, e.g., ][]{arevalo14,balokovic14,gandhi14,annuar15,bauer15,koss15,lansbury15,boorman16,puccetti16,ricci16}.

AGN obscuration is usually linked to the presence of a so-called ``dusty torus'', i.e., the gas and dust material surrounding the AGN accretion disk and responsible for the obscuration. This material is not continuously distributed, but more likely a clumpy distribution of optically thick dusty clouds \citep[e.g.,][]{elitzur06,risaliti07,honig07,nenkova08}. In the disk outflow scenario \citep{emmering92} the obscuration is caused by a clumpy disk wind and can be described using two main parameters: the AGN luminosity and Eddington ratio. 
This model predicts that in the low-luminosity regime (i.e., at bolometric luminosities L$_{\rm bol}\leq10^{42}$\,erg s$^{-1}$) the cloud outflow process cannot be sustained by the AGN and the broad line region (BLR) and the dusty clouds should both disappear \citep{nicastro00,elitzur09}. This prediction was observationally confirmed by \citet{burlon11}, which found a decrease in the fraction of obscured AGN detected by \swi\ at L$_{15-55\,keV}\leq10^{43}$\,erg s$^{-1}$. A similar behaviour may be linked to a change in the AGN accretion regime,  switching from  an optically thick, geometrically thin disk \citep[e.g.][]{shakura73} to a radiatively inefficient accretion flow \citep[e.g.,][]{narayan94,blandford99}. However, a recent work by \citet{brightman15} did not find evidence of decrease of the obscured AGN fraction at low luminosities. Consequently, the detection and study of low-luminosity, highly obscured AGN becomes strategic to better understand the physics of the torus and the accretion disk.

In this work, we analyze the 0.3-150 keV spectra of seven nearby Seyfert 2 galaxies detected with \swi\ in the 15--150 keV band and reported in the \swi\ 100-month catalog. These sources have L$_{15-55\,keV}\leq10^{43}$\,erg s$^{-1}$. We obtained a 10 ks follow-up with \textit{Chandra} ACIS-S in the 0.3--7 keV band for each object in the sample. The combination of \cha and \swi\ observations provides us an ideal dataset to properly characterize the main physical features of these sources, which are all expected to be heavily obscured and even Compton thick. In Section \ref{sec:sample} we describe our sample and the data reduction procedure adopted for the \cha data. In Section \ref{sec:results} we present the result of the spectral fitting procedure, a physical analysis of the best-fit models and a comparison with previous results for two of the seven sources. Finally, in Section 4 we discuss our results, focusing on the effectiveness of our selection procedure in detecting highly obscured AGN.

Throughout this work, we adopt a cosmology with $\Omega_m$= 0.27, $\Omega_\Lambda$=0.73, and H$_0$=71 km s$^{-1}$ Mpc$^{-1}$. Errors are quoted at 90\% confidence level unless otherwise stated.

\section{Sample properties and data reduction}\label{sec:sample}
The \textit{Swift} satellite \citep{gehrels04} is equipped with the wide-field (120$\times$90 deg$^2$) Burst Alert Telescope \citep[BAT; ][]{barthelmy05}. This telescope performs imaging in the 15-150 keV energy range and has been continuously observing the whole sky since its launch. The most recent catalog of sources detected in the BAT survey data is the 100-month catalog\footnote{http://bat.ifc.inaf.it/100m\_bat\_catalog/100m\_bat\_catalog\_v0.0.htm}, which contains sources detected down to a flux limit $f\sim$3.3 $\times$ 10$^{-12}$\,\flu\ in the 15-150 keV band. With its combination of good sensitivity and all-sky coverage, this instrument provides an excellent view of the hard X-ray low luminosity population in the nearby Universe. We point out that the 100-month BAT catalog is yet to be published (Segreto et al. in prep.), but all the data used in this work have been fully reprocessed.

To reprocess the BAT survey 100-month data available in the HEASARC public archive we used the BAT\_IMAGER code \citep{segreto10}. This software has been developed to analyze data from coded mask instruments and performs screening, mosaicking and source detection. The spectra used in this work are background subtracted and have been obtained averaging over the whole BAT exposure. We used the official BAT spectral redistribution matrix\footnote{http://heasarc.gsfc.nasa.gov/docs/heasarc/caldb/data/swift/\\bat/index.html}.

In Table \ref{tab:sample} we report the list of sources we analyze in this work. We selected from the \swi\ 100-month catalog seven nearby AGN ($z<$0.03), associated to Seyfert 2 galaxies, lacking of a ROSAT counterpart in the 0.1--2.4\,keV band.  
Two out of seven sources (ESO 116-G018 and NGC 5972) have never been observed before at 0.3--10 keV energies, and none of them has been previously observed using \textit{Chandra}.
The nearest source has $z$=0.0122 (i.e., $d\sim$52 Mpc), while the farthest has $z$=0.0297 ($d\sim$130 Mpc). The 15--55\,keV luminosity range spans from L$_{\rm BAT}\sim$ 2$\times$ 10$^{42}$\,erg s$^{-1}$ to L$_{\rm BAT}\sim$ 2 $\times$ 10$^{43}$\,erg s$^{-1}$. We selected low-redshift sources because the vast majority ($>$85\%) of \swi-detected CT-AGN have so far been discovered at $z<$0.04 \citep[see, e.g., ][]{burlon11,ricci15}. Moreover, the lacking of a ROSAT counterpart is already a hint of at least moderate obscuration, with N$_{\rm H}\gtrsim10^{22}$\,cm$^{-2}$ \citep[see, e.g., Figure 2 of][]{koss16}.

All the sources have been targeted with 10 ks \cha ACIS-S observations during \cha Cycle 17 (Proposal number 17900432, P.I. Marco Ajello). The data have been reduced with the CIAO \citep{fruscione06} 4.7 software and the \textit{Chandra} Calibration Data Base (\texttt{caldb}) 4.6.9, adopting standard procedures; no source shows significant pile-up, as measured by the CIAO \textsc{pileup\_map} tool. Source and background spectra have been extracted using the CIAO \texttt{specextract} tool. Source spectra have been extracted in circular regions of 4$^{\prime\prime}$, while background have been extracted from annuli having internal radius $r_{\rm int}$=7$^{\prime\prime}$ and external radius $r_{\rm out}$=20$^{\prime\prime}$, after a visual inspection to avoid contamination from nearby sources. Point-source aperture correction was applied during the source spectral extraction process. Finally, the spectra were binned in order to have at least 15 bins in the \cha spectrum: each bin contains from 8 to 20 counts, according to the source brightness.

\begingroup
\renewcommand*{\arraystretch}{1.2}
\begin{table*}
\centering
\scalebox{1.}{
\begin{tabular}{cccccccc}
\hline
\hline
\swi\ ID & Source name & R.A. & Dec & $z$ & L$_{\rm 15-55\,keV}$ &  Obs ID & Obs Date\\ 
& & deg (J2000) & deg (J2000) & & erg s$^{-1}$ & \\
\hline
J1253.5--4844 & NGC 4785 & 193.3639 & --48.7492 & 0.0123 & 2.3 $\times$ 10$^{42}$ & 18074 & 2016 Apr 16\\ 
J1024.6--2332 & ESO 500-G034  & 156.1310 & --23.5530 & 0.0122 & 1.8 $\times$ 10$^{42}$ & 18075 & 2016 Feb 08\\ 
J0333.7--0504 & NGC 1358 & 53.4153 & --5.0894 & 0.0134 & 2.0 $\times$ 10$^{42}$  & 18076 & 2015 Nov 21\\ 
J0324.8--6043 & ESO 116-G018 & 51.2210 & --60.7384 & 0.0185 & 2.5 $\times$ 10$^{42}$ & 18077 & 2016 Aug 05\\ 
J0448.9--5739 & ESO 119-G008 & 72.2364 & --57.6592 & 0.0229 & 4.4 $\times$ 10$^{42}$ & 18078 & 2016 Jul 09\\ 
J2234.9--2543 & ESO 533-G050 & 338.7076 & --25.6769 & 0.0264 & 1.1 $\times$ 10$^{43}$ & 18079 & 2016 May 22\\ 
J1539.0+1701 & NGC 5972 & 234.7257 & 17.0262 & 0.0297 & 1.0 $\times$ 10$^{43}$ & 18080 & 2016 Apr 04\\ 
\hline
\hline
\end{tabular}}\caption{\normalsize Summary of the sources in our sample. All the \swi\ IDs come from the Palermo BAT Catalog fourth version (4PBC). ``Obs ID'' and ``Obs Date'' refer to the \cha\ observations.} 
\label{tab:sample}
\end{table*}
\endgroup

\section{Spectral fitting results}\label{sec:results}
In Table \ref{tab:results_mytorus} we report the results of the spectral fitting of the seven \swi\ Seyfert 2 galaxies in our sample. The spectral fitting has been performed using XSPEC v. 12.9.0 \citep{arnaud96}. Galactic absorption was determined using the Heasoft tool \texttt{nh} \citep{kalberla05}. 

All sources have first been fitted with a basic power law with photon index $\Gamma$ and intrinsic absorption $N_{\rm H, z}$ caused by the dusty torus surrounding the AGN accretion disk. The intrinsic absorption is computed using the \texttt{MyTorus} model \citep{murphy09}, which assumes a toroidal distribution of the absorbing material and is particularly effective in fitting spectra of heavily obscured objects. 

We determined that for all seven sources in the sample more complex models than the basic absorbed power law one were required to significantly improve the fit. The significance of the additional components has been verified performing a F-test, when statistically allowed; in the remaining cases \citep[i.e., when a Gaussian or a reflection component were added, see ][]{protassov02} the fits without the additional components always had reduced $\chi^2$, $\chi^2$/(degrees of freedom)$>$1.5, while the best-fit with the additional component has reduced $\chi^2$ $\sim$1. 

These are the different components we used in the fit, in addition to the basic absorbed power law. The best-fit models for each source are reported in Table \ref{tab:results_mytorus}.
\begin{enumerate}
\item Six spectra require a second power law component, having photon index $\Gamma_2$=$\Gamma_1$, the photon index of the main power law, and not affected by the intrinsic absorption $N_{\rm H, z}$. This second power law accounts for a fraction of AGN emission unabsorbed by the torus and/or for a scattered component, i.e., light deflected rather than absorbed by the dust and gas. The normalization of this second component is always $<$1\% than the normalization of the main power law.
\item Four fits are significantly improved adding the \texttt{MyTorus} fluorescent emission line component to model an excess in the spectrum at energy $E\sim$6.4\,keV. The line in the model was fixed at 6.4 keV (rest-frame) and accounts for the presence of the iron K$\alpha$ line. 
The iron K$\alpha$ equivalent width varies in the range EW=[0.15--0.75] keV (we report the EW values in Table \ref{tab:results_mytorus}).
\item Two sources in the sample require the addition of a black body component with $kT<$0.3 keV to model the characteristic ``soft excess'' commonly observed in the AGN spectra and possibly due to the presence of warm gas close to the AGN accretion disk or in the BLR \citep[see, e.g., ][]{risaliti04}. In order to understand the origin of this excess, for these two sources (NGC 1358 and NGC 5972) we used the equations reported in \citet{lehmer10} to quantify the expected contribution of star-forming (SF) processes to the X-ray emission. We used the IRAS 8--1000\,$\mu$m luminosities to compute the star-formation rate (SFR) and SFR to estimate L$^{\rm SF}_{\rm 2-10\,keV}$. For both sources, we find L$^{\rm SF}_{\rm 2-10\,keV}\leq$10$^{39}$ \lu, i.e., $\lesssim$0.1\% of the AGN luminosity in the energy range, therefore safely ruling out a significant SF contribution to the X-ray emission.
\item One source in the sample (NGC 5972) shows an excess in the \swi\ spectrum at E$\sim$30 keV which is best-fitted with the \texttt{PEXRAV} model \citep{magdziarz95}, which parametrizes a power law with a significant reflection component. This reflection component is caused by cold material surrounding the accretion disk. Due to the low statistics of the spectrum of NGC 5972, we fixed the reflection scaling factor $R$ to 1. Since it is $R=\Omega/2\pi$, where $\Omega$ is the solid angle of the cold material, responsible for the reflection, visible from the hot corona, responsible from the X-ray emission, $R$=1 is the case where the reflection is produced by an infinite slab isotropically illuminated by the corona emission.
\end{enumerate}

Since the \swi\ spectrum is averaged on 100 months of observations, while the \cha one is obtained by a single observation, the joint fit of the \cha and \swi\ spectra requires the addition of a multiplicative constant ($C_{\rm BAT}$) to the \swi\ spectrum, to account for both flux variability and cross-correlation offsets. We examined the \swi\ light curves of the seven sources in our sample and none of them shows strong flares, the flux variability being limited to 20--40\%. For five out of seven sources, the best-fit constant we obtained is consistent (within the errors) to $C_{\rm BAT}$=1 and can be explained with the variability observed in the \swi\ light curves. For two AGN (NGC 1358 and ESO 116-G018), however, the best-fit constant we find is significantly higher, being $\sim$9 and $\sim$3, respectively. Such a high variability cannot being ruled out on the basis of our observations, since NGC 1358 and/or ESO 116-G018 could be changing-look AGN (i.e., sources with a significant change in flux and/or amount of obscuring material); however, similar flux variations are extremely unlikely in highly obscured sources, the reflection component washing out fluctuations on short timescales \citep[see, e.g.,][]{gandhi15}. Therefore, for these two sources we fixed $C_{\rm BAT}$ to 1.

\begin{figure}
  \centering
  \includegraphics[width=1.\linewidth]{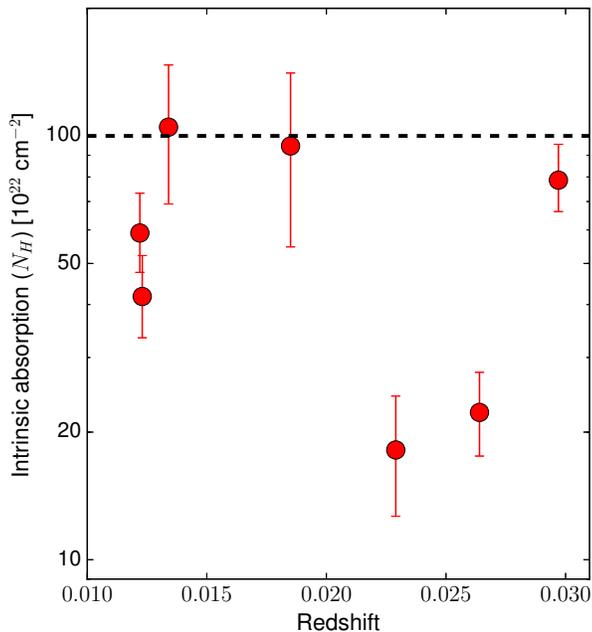}
\caption{\normalsize Intrinsic absorption $N_{\rm H, z}$, computed assuming $\theta$=90\degree, as a function of redshift, for the seven sources in our sample. The horizontal dashed line marks the Compton-thick regime, i.e., $N_{\rm H, z}\geq$10$^{24}$\,cm$^{-2}$.}\label{fig:z_vs_nh}
\end{figure}

\begin{figure*}
\begin{minipage}[b]{.5\textwidth}
  \centering
  \includegraphics[width=1.02\textwidth]{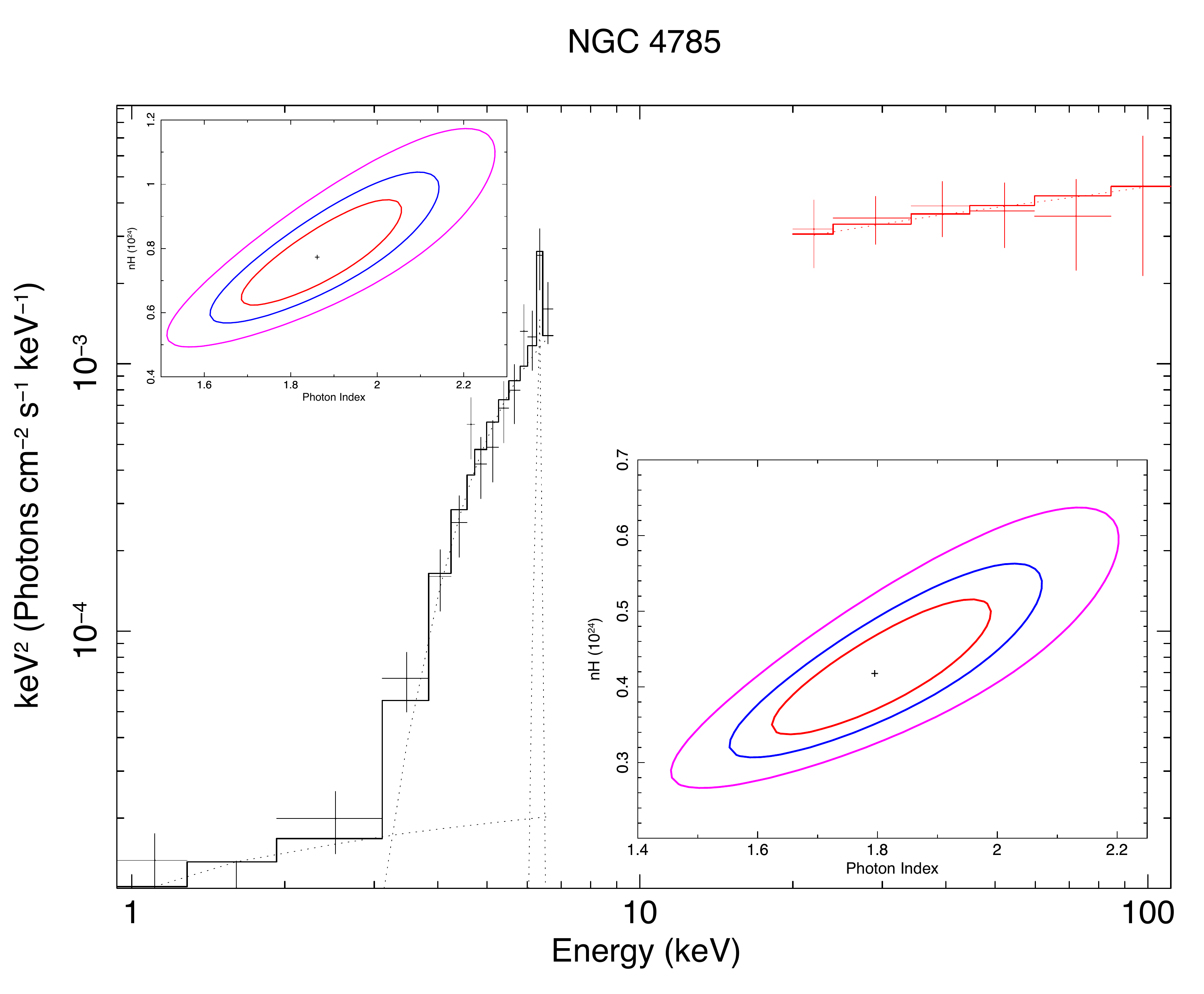}
  \end{minipage}
\begin{minipage}[b]{.5\textwidth}
  \centering
  \includegraphics[width=1.02\textwidth]{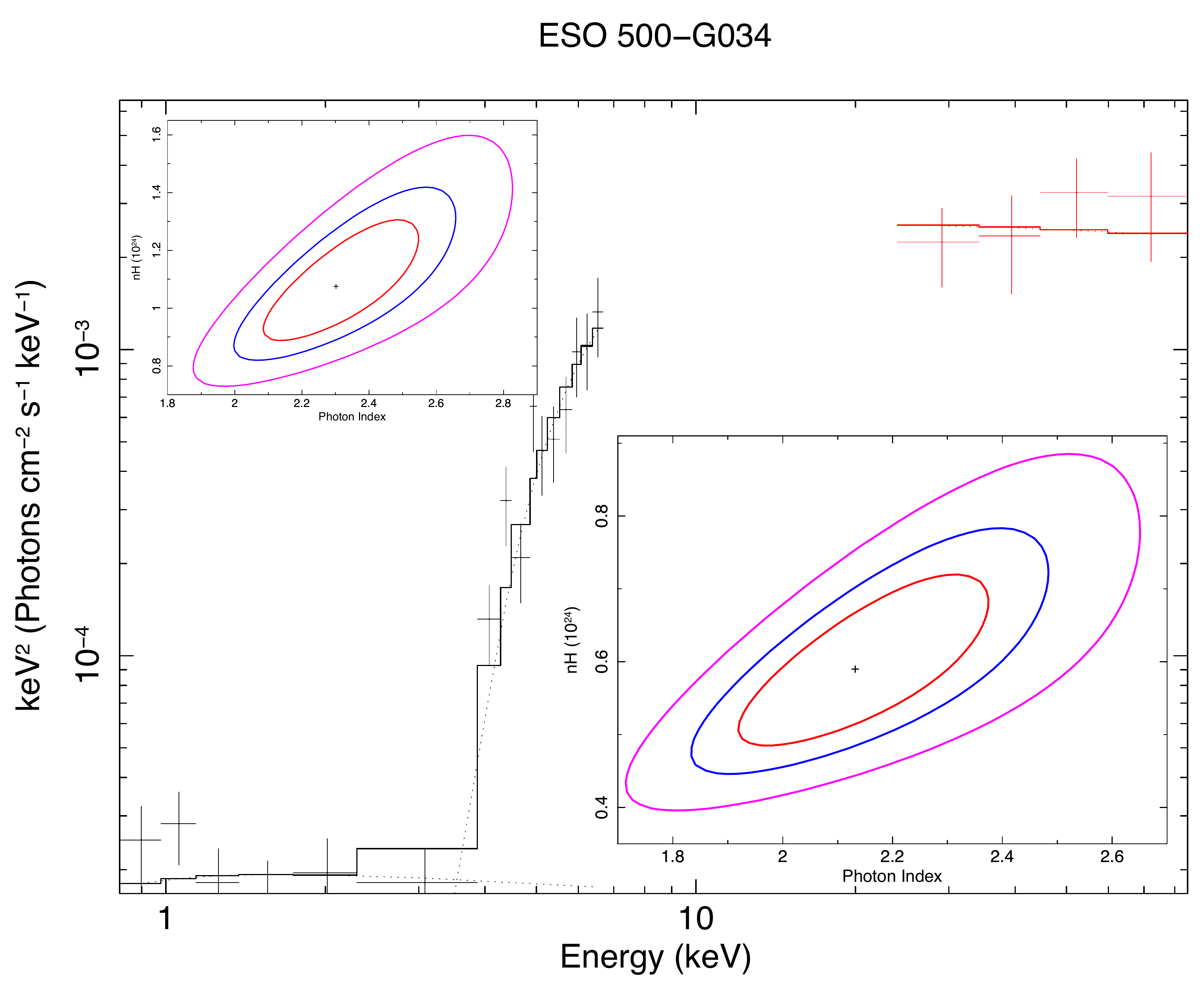}
  \end{minipage}
\begin{minipage}[b]{.5\textwidth}
  \centering
  \includegraphics[width=1.02\textwidth]{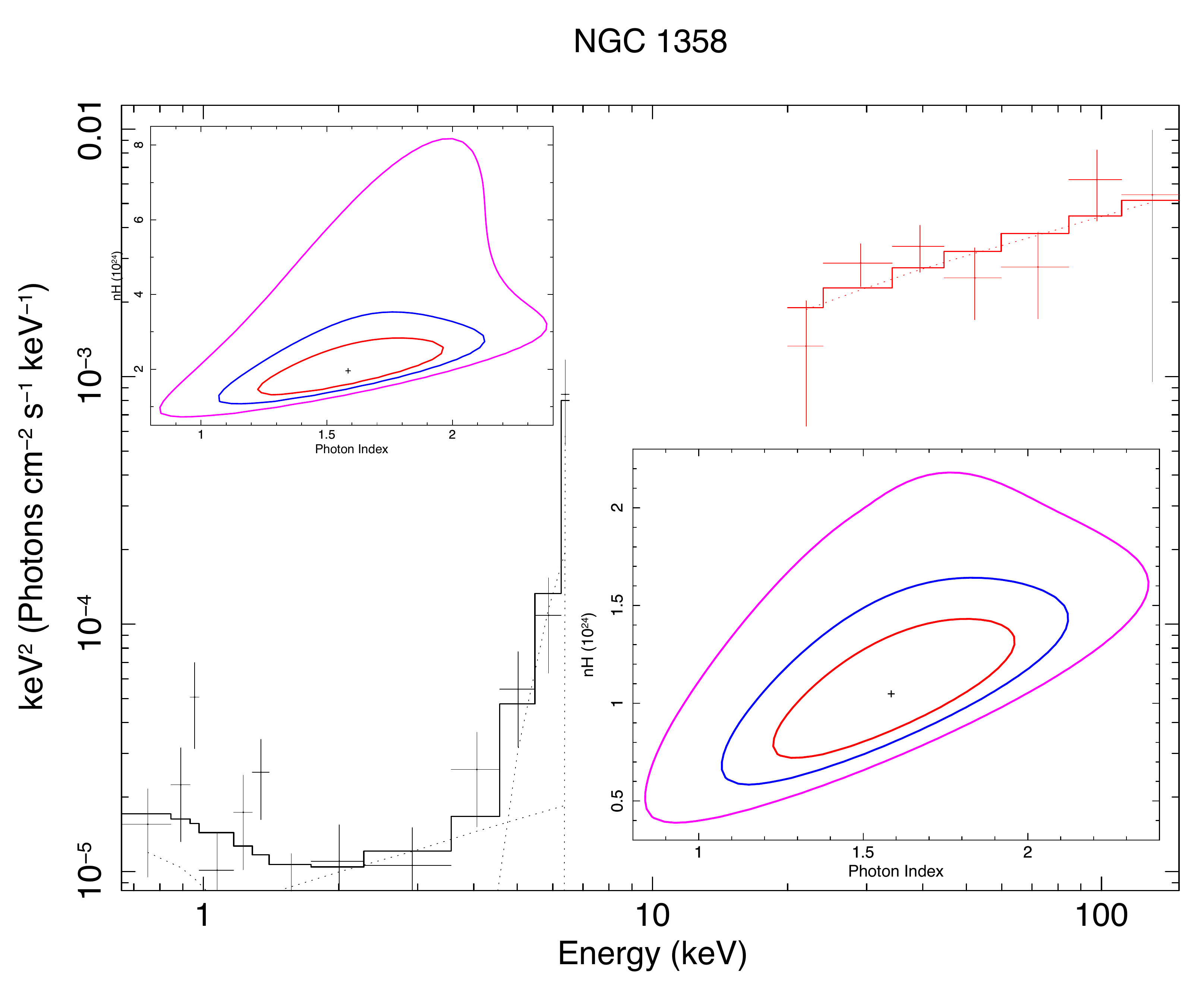}
  \end{minipage}
  \begin{minipage}[b]{.5\textwidth}
   \centering
  \includegraphics[width=1.02\linewidth]{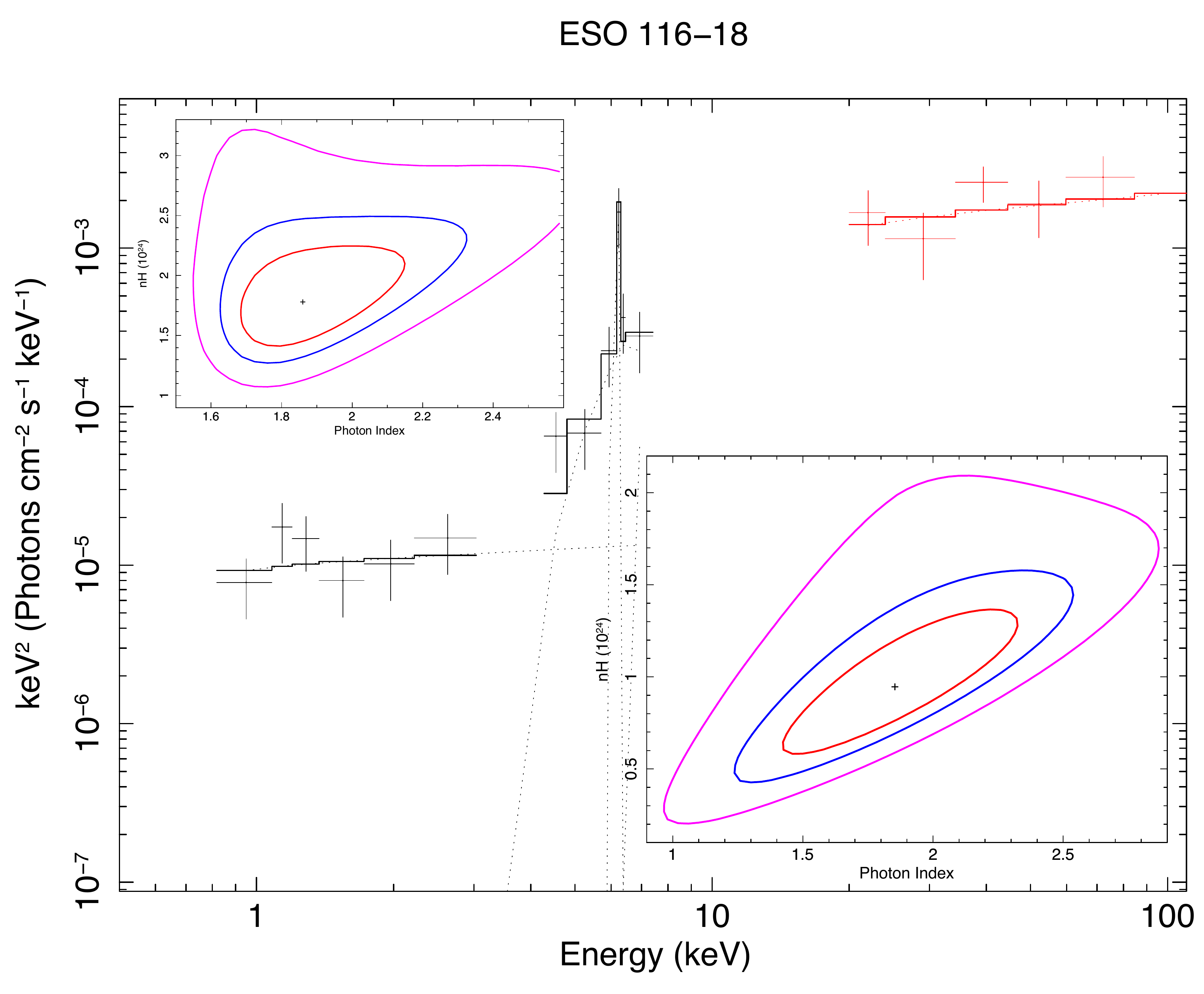}
  \end{minipage}
   \begin{minipage}[b]{.5\textwidth}
   \centering
  \includegraphics[width=1.02\linewidth]{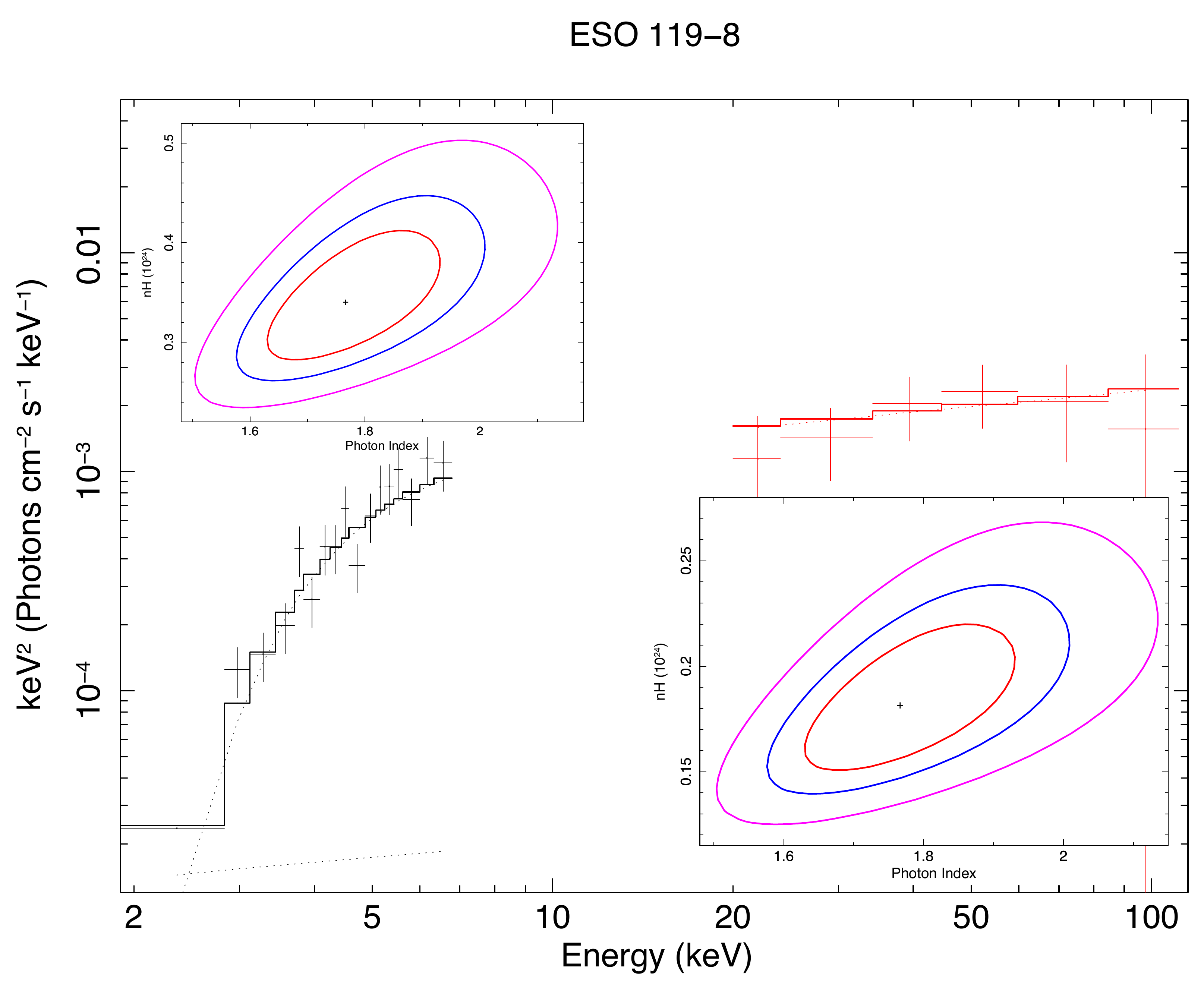}
  \end{minipage}
   \begin{minipage}[b]{.5\textwidth}
   \centering
  \includegraphics[width=1.02\linewidth]{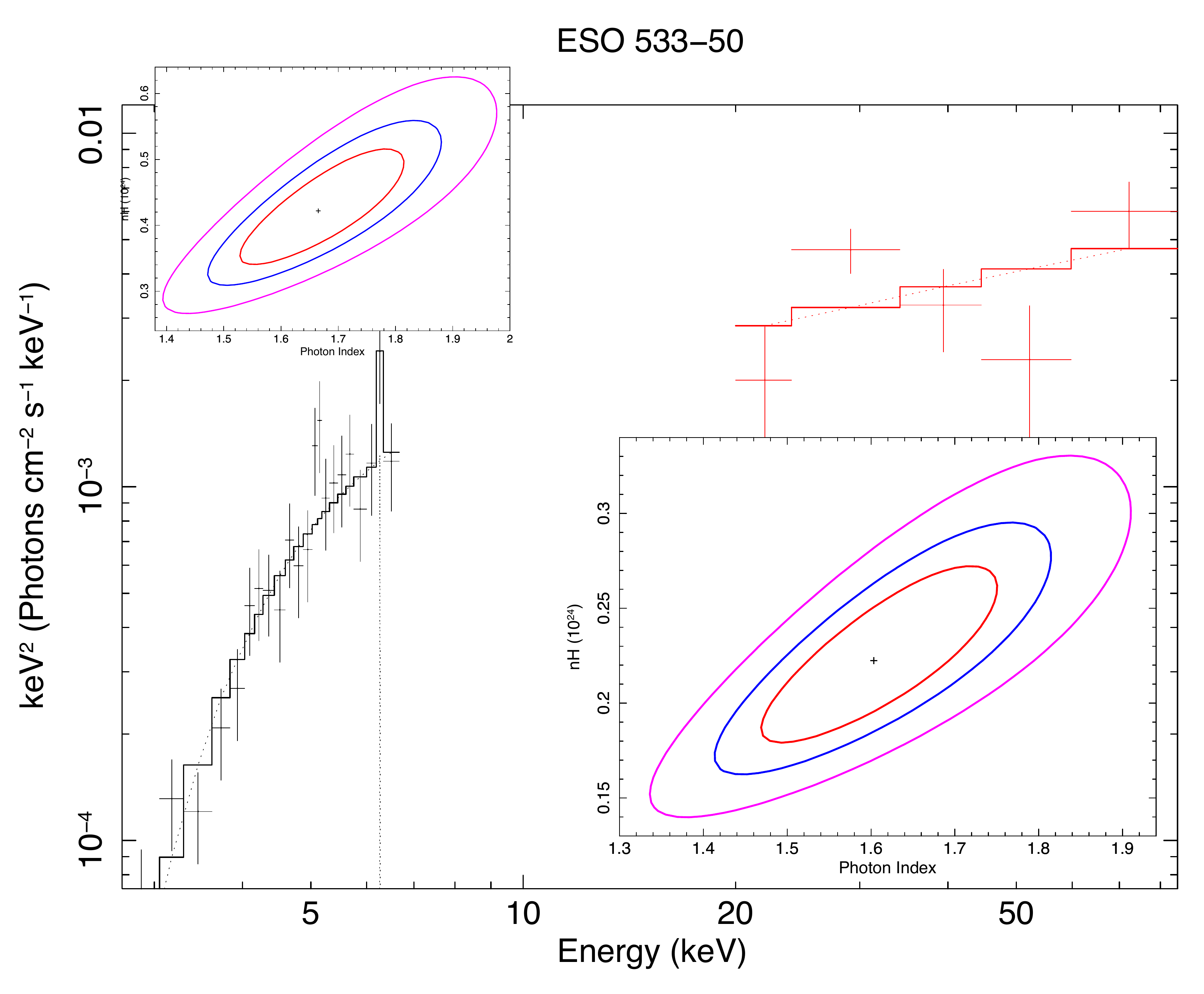}
  \end{minipage}
\caption{\normalsize \cha (black) and \swi\ (red) spectra of six of the seven sources in our sample. The best-fitting model is plotted as a solid line, while the single components are plotted as dotted lines. In the inset, the confidence contours at 68, 90 and 99\% confidence level for $\Gamma$ and $N_{\rm H, z}$ (in units of 10$^{24}$ cm$^{-2}$) are shown, assuming $\theta$=90\degree\ (bottom right) and $\theta$=65\degree\ (top left).}\label{fig:spec_contour_1}
\end{figure*}

\begin{figure*}
  \centering
  \includegraphics[width=0.6\linewidth]{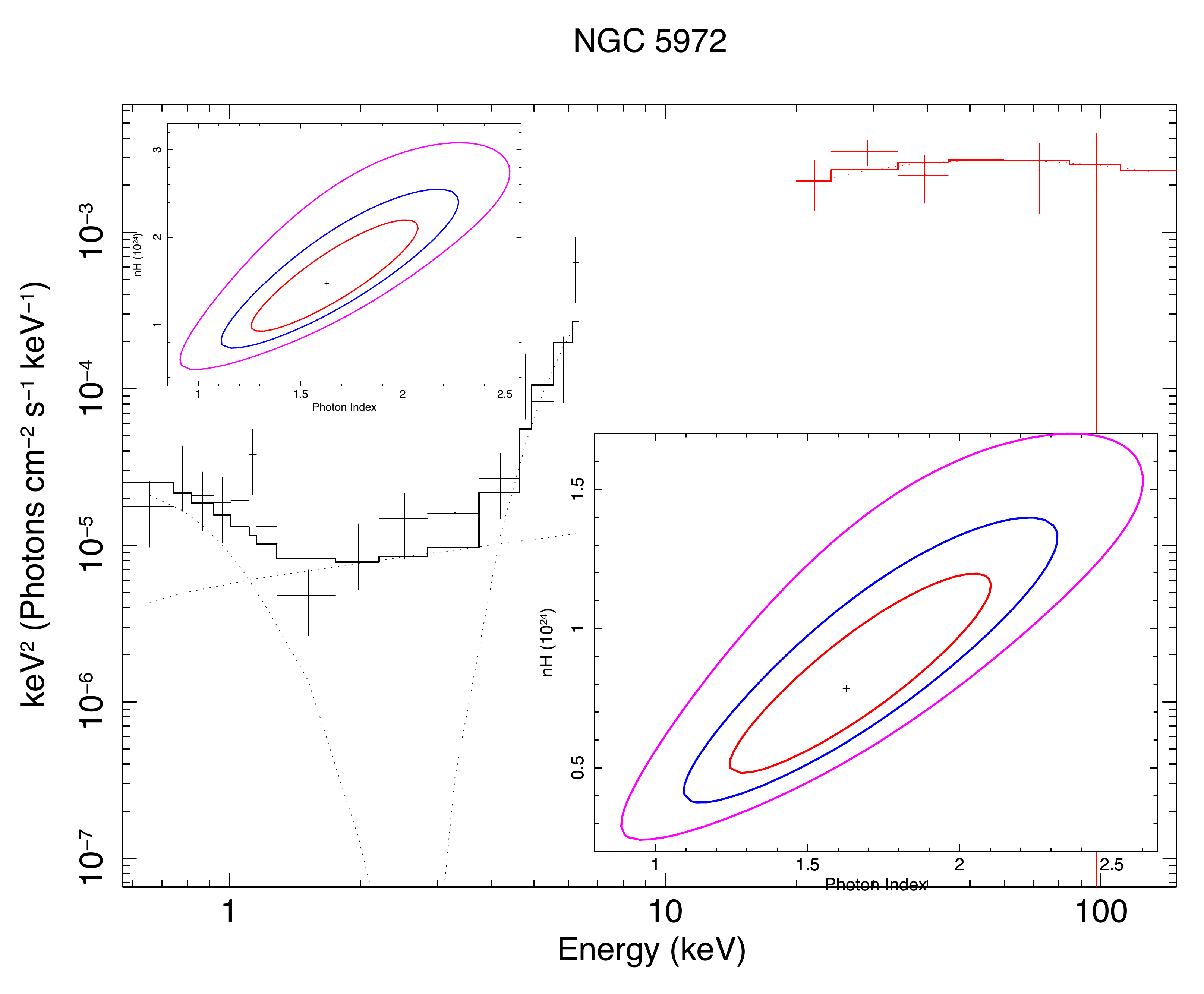}
\caption{\normalsize \cha (black) and \swi\ (red) spectra of NGC 5972. The best-fitting model is plotted as a solid line, while the single components are plotted as dotted lines. In the inset, the confidence contours at 68, 90 and 99\% confidence level for $\Gamma$ and $N_{\rm H, z}$ (in units of 10$^{24}$ cm$^{-2}$) are shown, assuming $\theta$=90\degree\ (bottom right) and $\theta$=65\degree\ (top left).}\label{fig:spec_contour_2}
\end{figure*}

In Figure \ref{fig:z_vs_nh} we show the $N_{\rm H, z}$ distribution versus redshift for the sources in our sample, while in Figures \ref{fig:spec_contour_1} and \ref{fig:spec_contour_2} we report the seven spectra and their best fits. In the inset, we also report the confidence contours for $\Gamma$ and $N_{\rm H, z}$: as can be seen, all the sources are heavily obscured. We will discuss in detail the $N_{\rm H, z}$ measurements in the next section: here we point out that, as a consequence of the heavy obscuration, which affects mainly the 0.3--7\,kev band, the ratio between the flux in the 15--150\,keV band and the flux in the 0.3--7\, keV one is $>$10 for all the sources in our sample and even $>$30 for the three most obscured objects.

\begingroup
\renewcommand*{\arraystretch}{1.5}
\begin{table*}
\centering
\scalebox{0.87}{
\begin{tabular}{ccccccccccc}
\hline
\hline
\swi\ ID & $N_{\rm H, gal}$ & $N_{\rm H, z,\theta=90}$ & $N_{\rm H, z,\theta=65}$ & $\theta_{max}$ & $\Gamma$ & EW & $kT_{BB}$ & SC & $\chi^2$/DOF & Model\\
& 10$^{20}$ cm$^{-2}$ & 10$^{22}$ cm$^{-2}$ & 10$^{22}$ cm$^{-2}$ & \degree & & keV & keV & \\
\hline
NGC 4785 & 12.2 & 41.38$_{-8.75}^{+10.94}$ & 77.37$_{-16.47}^{+20.65}$ & -- & 1.86$_{-0.19}^{+0.21}$ & 0.48$_{-0.34}^{+0.33}$  & -- & 0.47$\pm$0.13 & 7.9/15 & \texttt{po+MyT*(po+zga)}\\ 
ESO 500-G034 & 5.1 & 57.41$_{-10.80}^{+13.48}$ & 107.60$_{-20.24}^{+25.26}$ & 66 & 2.30$_{-0.24}^{+0.27}$ & -- & -- & 0.18$\pm$0.06 & 12.3/15 & \texttt{po+MyT*po}\\ 
NGC 1358 & 3.8 & 104.87$_{-35.80}^{+42.14}$ & 196.44$_{-67.03}^{+211.09}$ & 90 &1.59$_{-0.45}^{+0.46}$ & 0.75$_{-0.74}^{+0.75}$ & 0.16$_{-0.11}^{+0.12}$ & 0.77$\pm$0.42 & 14.8/13 & \texttt{po+MyT*(bb+po+zga)}\\
ESO 116-G018 & 3.1 & 94.68$_{-39.97}^{+46.06}$ & 178.13$_{-75.34}^{+86.19}$ & 81 & 1.86$_{-0.48}^{+0.51}$ & 1.41$_{-0.94}^{+2.67}$ & -- & 0.52$\pm$0.29 & 17.2/14 & \texttt{po+MyT*(po+zga)}\\ 
ESO 119-G008 & 1.3 & 18.15$_{-5.49}^{+6.18}$ & 34.02$_{-10.28}^{11.57}$ & -- & 1.77$_{-0.18}^{+0.21}$ & -- & -- & 0.39$\pm$0.10 & 17.7/20 & \texttt{po+MyT*po}\\ 
ESO 533-G050 & 1.5 & 22.56$_{-4.74}^{+5.45}$ & 42.24$_{-8.88}^{+10.23}$ & -- & 1.65$_{-0.14}^{+0.18}$ & 0.15$_{-0.15}^{+0.17}$ & -- & 0.54$\pm$0.12 & 25.6/22 & \texttt{MyT*(po+zga)}\\ 
NGC 5972 & 3.0 & 78.62$_{-12.34}^{+16.94}$ & 147.27$_{-59.22}^{+79.89}$ & 72 & 1.63$_{-0.09}^{+0.23}$ & -- & 0.14$_{-0.05}^{+0.07}$ & 0.65$\pm$0.35 & 17.6/16 & \texttt{po+MyT*(bb+pexr)}\\ 
\hline
\hline
\end{tabular}}\caption{\normalsize Best fit properties for the seven Seyfert 2 galaxies in our sample. The XSPEC components reported in the ``model'' columns are the following: \texttt{pow} is a power law with photon index $\Gamma$; when two power laws are present, their photon index is the same. \texttt{zga} is the \texttt{MyTorus} Iron $K\alpha$ component with equivalent width EW. \texttt{bb} is a black-body component having temperature $kT_{BB}$ and fitting a soft excess observed below 1 keV. \texttt{pexr} is a power law with a cold matter reflection component. The intrinsic absorption $N_{\rm H, z}$ is modeled with \texttt{MyTorus}  (\texttt{MyT} in the table), with inclination angle $\theta$ equal to 65\degree\ or to 90\degree. $\theta_{max}$ is the maximum inclination angle for which is $N_{\rm H, z}>$10$^{24}$\,cm$^{-2}$. SC is the spectral curvature value computed following \citet{koss16}}\label{tab:results_mytorus}
\end{table*}
\endgroup

\subsection{Intrinsic absorption and its trend with torus inclination angle}\label{sec:nh}
One of the \texttt{MyTorus} model parameters is $\theta$, the inclination angle between the observer line of sight and the torus symmetry axis, with $\theta$=0\degree\ being a face-on observing angle and $\theta$=90\degree\ being an edge-on observing angle. $\theta$=60\degree\ is the torus opening angle, while $\theta$=61\degree\ is the lowest angle for which the observer line of sight intercepts the torus.

The quality of our spectra does not allow us to constrain $N_{\rm H, z}$ and $\theta$ simultaneously. Therefore, we fitted our data twice: the first one assuming $\theta$=90\degree, i.e., the edge-on scenario, the second one with $\theta$=65\degree, i.e., an angle closer to the angle for which the observer line of sight intercepts the torus. As can be seen in Table \ref{tab:results_mytorus}, assuming $\theta$=65\degree\ implies obtaining best-fit $N_{\rm H, z}$ values $\sim$2 times higher than those obtained assuming $\theta$=90\degree.

As already pointed out in the previous section, all sources in our sample are heavily obscured. Even in the worst case scenario, i.e., assuming that all sources are observed edge-on, ($\theta$=90\degree), all the objects have $N_{\rm H, z}>$10$^{23}$\,cm$^{-2}$ at a 99\% confidence level. Moreover, four out of seven sources can in principle be CT-AGN, assuming that we are observing them with a $\theta$ close to the torus opening angle: in such a scenario, the path of the photons across the torus and the intrinsic absorption value are respectively smaller and larger than in the case with $\theta$=90\degree.

To complete our analysis, we re-fitted the spectra with two other models to describe the absorbed power-law: the first one is the standard photo-electric absorption (\texttt{zwabs} in XSpec), while the other is \texttt{pclabs}, which describes the X-ray emission of an isotropic source of photons, corrected by the absorption caused by a spherical distribution of material. In this second model, Compton scattering is taken into account. In both cases, we find that the $N_{\rm H, z}$ best-fit values are consistent (within the errors) with the \texttt{MyTorus} best-fit values obtained assuming $\theta$=90\degree. These results may indicate that reliable $\theta$ values are closer to the $\theta$=90\degree\ limit than to the $\theta$=65\degree\ one. However, a proper disentanglement of the $N_{\rm H, z}$--$\theta$ can be broken only increasing the 0.3--7\,keV band exposure (see Section \ref{sec:discuss} for further details).

\subsection{Comparison with previous results}\label{sec:compare}
Two sources in our sample have already been observed in the 0.5--10\,keV band. In this section, we present previously published works and we compare their results with ours.

NGC 1358 was targeted with a 10\,ks \xmm\ observation in 2005 (PI: I. Georgantopoulos): the spectral fitting results for this observation (the \swi\ spectrum was not used in this work) have been published in \citet{marinucci12}. Their results are in good agreement with ours: they find that NGC 1358 is consistent with being a CT-AGN, although with large uncertainties, having best-fit intrinsic absorption $N_{\rm H, z}$=1.3$^{+8.5}_{-0.6}\times$10$^{24}$ cm$^{-2}$.

NGC 4785 has already been studied by \citet{gandhi15}: they combined the \swi\ observation used in this work with a 79\,ks \textit{Suzaku} observation. They fitted the data with \texttt{PEXRAV}, with \texttt{MyTorus} and with the \texttt{TORUS} model by \citet{brightman11}. In all cases, they found the source being Compton-thick, the measured intrinsic absorption varying from $N_{\rm H, z}$=1.4$^{+0.5}_{-0.9}$ cm$^{-2}$ while fitting the data with \texttt{PEXRAV} to $N_{\rm H, z}$=2.7$^{+3.8}_{-0.8}$ cm$^{-2}$ fitting the data with \texttt{TORUS}. Our  $N_{\rm H, z}$ estimate is therefore consistent, within the errors, with their \texttt{PEXRAV} best-fit, and lie below their results with \texttt{MyTorus} and \texttt{TORUS}. While the hypothesis that this discrepancy can be partially driven by the low statistics in our fit, which has only 16 degrees of freedom, compared to the 89 of the \citet{gandhi15} one,  it is also possible that the change in the measure intrinsic absorption  is due to intrinsic variability in the amount of obscuring material in the time between the \textit{Suzaku} observation (taken in July 2013) and the \cha one (taken in April 2016). A significant change of the AGN luminosity seems less likely, since the power-law photon index we measured ($\Gamma$=1.80$_{-0.19}^{+0.21}$) is in good agreement with the one of \citet{gandhi15} ($\Gamma$=2.1$_{-0.3}^{+0.4}$ using \texttt{MyTorus}). Finally, \citet{gandhi15} find evidence of a strong Iron K$\alpha$ line in NGC 4785, with EW=1.1$^{+u}_{-0.4}$ keV (where $u$ means unconstrained), while the EW of the line in our fit is weaker (EW=0.48$_{-0.34}^{+0.33}$). This change in line intensity may strenghten the hypothesis that we are observing a variation in the amount of obscuring material, since CT-AGN usually show prominent iron lines with EW$>$1\,keV \citep[see, e.g.,][]{koss16}.

\section{Discussion and conclusions}\label{sec:discuss}
In this work, we presented the spectral fitting over the 0.3--150 keV energy range, obtained combining \cha and \swi\ observations, of seven AGN at $z<$0.03 selected from the 100-month \swi\ all-sky survey catalog. These sources are associated to Seyfert 2 galaxies and lack of a ROSAT counterpart in the 0.1--2.4\,keV band. The combined \cha and \swi\ spectra have then been fitted with different models, to estimate spectral parameters such as the spectral photon index $\Gamma$ and the intrinsic absorption $N_{\rm H, z}$.

As discussed in Section \ref{sec:nh}, all the sources in the sample are heavily obscured: even in the (unlikely) scenario of all the sources being edge-on (i.e., with inclination angle $\theta$=90\degree) the less obscured object (ESO 119-G008) has $N_{\rm H, z}$=1.82$_{-0.55}^{+0.62}$ $\times$ 10$^{23}$ cm$^{-2}$ and all sources have $N_{\rm H, z}>$ 10$^{23}$ cm$^{-2}$ at a 99\% confidence level. Moreover, one source (NGC 1358) has best-fit intrinsic absorption $N_{\rm H, z}$=1.05$_{-0.36}^{+0.42}$ $\times$ 10$^{24}$ cm$^{-2}$, i.e., consistent with being a Compton thick AGN. 

Furthermore, we performed a further fitting analysis varying $\theta$ in 1\degree\ steps, and we found  that five of the sources in our sample can be CT-AGN, if the inclination angle $\theta$ between the observer line of sight and the torus rotation axis is small enough (see column $\theta_{max}$ in Table \ref{tab:results_mytorus}). 

A first important result of this work is therefore the finding of an effective way to detect obscured AGN in the nearby Universe, taking advantage of the combination of low limiting flux and wide field of view of the \swi\ all-sky survey. It is worth noticing that, while with our selection technique we detect a population of sources with $N_{\rm H, z}>$ 10$^{23}$ cm$^{-2}$, the observed fraction of unobscured AGN (i.e., with $N_{\rm H, z}<$ 10$^{22}$ cm$^{-2}$) obtained without applying any selection criteria is 35--55\% \citep{burlon11}.

\citet{koss16} developed a technique to identify CT-AGN in AGN with low counts statistics observed with \swi\ or \nustar, finding that the fraction of CT-AGN at $z<$0.03 is $\sim$22\%. The \citet{koss16} method is based on the curvature of the AGN spectrum between 14 and 50 keV, parametrized as follows:
\begin{equation}
SC_{BAT}=\frac{-3.42A-0.82B+1.65C+3.58D}{Tot},
\end{equation} 

where $A$, $B$, $C$, $D$ and $Tot$ are the count rates measured with \swi\ in the 14-20\,keV, 20-24\,keV, 24-35\,keV, 35-50\,keV and 14-50\,keV, respectively. $SC_{BAT}$=0.4 is the CT-AGN selection threshold: in \citet{koss16} work, seven of the nine sources with $SC_{BAT}>$0.4 in their sample have $N_{\rm H, z}>$ 10$^{24}$ cm$^{-2}$, and the remaining two are significantly obscured ($N_{\rm H, z}>$ 5 $\times$ 10$^{23}$ cm$^{-2}$).

In Table \ref{tab:results_mytorus} we report the $SC_{BAT}$ values for the seven sources in our sample: as can be seen, the \citet{koss16} assumption on $SC_{BAT}$ is supported by our results. In fact, the most obscured source in our sample (NGC 1358) has $SC_{BAT}$=0.77$\pm$0.42, i.e., the highest $SC_{BAT}$ value for the sources in our sample, although the uncertainty on the SC value is quite large. Other two highly obscured sources (ESO 116-G018 and NGC 5972), both have $SC_{BAT}>$0.4 ($SC_{BAT}$=0.52$\pm$0.29 and $SC_{BAT}$=0.65$\pm$0.35, respectively), although both the sources have $SC_{BAT}<$0.4 at a 90\% confidence level. These two objects have $N_{\rm H, z}<$10$^{24}$ cm$^{-2}$ assuming $\theta$=90\degree, but may have $N_{\rm H, z}>$10$^{24}$ cm$^{-2}$ assuming reasonable values of $\theta$ (81\degree\ and 72\degree, respectively).
Finally, two sources with $SC_{BAT}>$0.4, and therefore candidate CT-AGN according to the \citet{koss16} equation, have $N_{\rm H, z}\sim$ 2--4 $\times$ 10$^{23}$ cm$^{-2}$ assuming $\theta$=90\degree.

The results of this work represent a first step in the process of identifying and analyzing CT-AGN in the local Universe. To improve our findings, we plan to extend our analysis in two ways:
\begin{enumerate}
\item Increasing the exposure for the objects we described in this work, using \xmm, \cha and \textit{NuSTAR}. The 0.3-7 keV spectra studied in this work have been obtained with observations of only 10\,ks each, and have between 80 (for the two most obscured AGN in our sample) and 270 (ESO 533-G050) net counts in the 0.3-7 keV band: breaking the degeneracy between the intrinsic absorption $N_{\rm H, z}$ and the inclination angle $\theta$, therefore definitely assessing if a source is or is not a CT-AGN, requires a larger counts statistics. 
As a comparison, \citet{tanimoto16} constrained simultaneously $N_{\rm H, z}$ and $\theta$ for the three candidate CT-AGN observed with \textit{Suzaku} for 30-70\,ks and having each about 500 net counts in the 0.3-7\,keV band.
\item Increasing the sample of candidate CT-AGN. The selection criteria we adopted in this work proved very effective, and the \swi\ 100-month catalog contains other $\sim$5--10 sources at $z<$0.04 that are Seyfert 2 galaxies lacking of ROSAT counterpart and which have not been observed by either \cha or \xmm. We plan to target these sources with future \cha and \xmm\ observations.
\end{enumerate}

\bibliographystyle{aa}
\bibliography{ctagn_paper_arxiv}


\end{document}